\documentclass[aps,amsmath,amssymb,prl,preprint,preprintnumbers,superscriptaddress,showpacs,showkeys,lengthcheck]{revtex4}

\usepackage{graphicx}
\usepackage{dcolumn}
\usepackage{bm}
\usepackage{hyperref}

\newcommand{\bscco}{$\textrm{Bi}_2\textrm{Sr}_2\textrm{CaCu}_2\textrm{O}_{8+\delta}$}

\begin{document}


\title{Nanoscale Proximity Effect in the High Temperature Superconductor \bscco}


\author{Colin V. Parker}
\affiliation{Joseph Henry Laboratories and Department of Physics, Princeton University, Princeton, NJ 08544, USA}
\author{Aakash Pushp}
\altaffiliation{Department of Physics, University of Illinois at Urbana-Champaign, Urbana, IL 61801, USA}
\affiliation{Joseph Henry Laboratories and Department of Physics, Princeton University, Princeton, NJ 08544, USA}
\author{Abhay N. Pasupathy}
\affiliation{Joseph Henry Laboratories and Department of Physics, Princeton University, Princeton, NJ 08544, USA}
\author{Kenjiro K. Gomes}
\altaffiliation{Current Address: Department of Physics, Stanford University, Stanford, CA 94305, USA}
\affiliation{Joseph Henry Laboratories and Department of Physics, Princeton University, Princeton, NJ 08544, USA}
\author{Jinsheng Wen}
\author{Zhijun Xu}
\affiliation{Condensed Matter Physics and Materials Science, Brookhaven National Laboratory (BNL), Upton, NY 11973, USA}
\author{Shimpei Ono}
\affiliation{Central Research Institute of Electric Power Industry, Komae, Tokyo 201-8511, Japan}
\author{Genda Gu}
\affiliation{Condensed Matter Physics and Materials Science, Brookhaven National Laboratory (BNL), Upton, NY 11973, USA}

\author{Ali Yazdani}
\affiliation{Joseph Henry Laboratories and Department of Physics, Princeton University, Princeton, NJ 08544, USA}
\email{yazdani@princeton.edu}


\date{\today}

\begin{abstract}
High temperature cuprate superconductors exhibit extremely local nanoscale phenomena and strong sensitivity to doping. While other experiments have looked at nanoscale interfaces between layers of different dopings, we focus on the interplay between naturally inhomogeneous nanoscale regions. Using scanning tunneling microscopy to carefully track the same region of the sample as a function of temperature, we show that regions with weak superconductivity can persist to elevated temperatures if bordered by regions of strong superconductivity. This suggests that it may be possible to increase the maximum possible transition temperature by controlling the distribution of dopants.
\end{abstract}


\maketitle


Many properties of high temperature cuprate superconductors,
such as the transition temperature $T_c$, and the presence of pseudogap,
have a strong doping dependence. What happens when two regions with
different doping are put into contact? Since pairing in
cuprate superconductors has a very local
character\cite{Gomes:2007p301,Howald:2001p505,Pan:2001p393,McElroy:2005p326},
one expects the coherence length to be short, and any proximity effects to
occur only on microscopic length scales. Nonetheless, some experiments
have demonstrated that high-$T_c$ Josephson junctions can be made with thicknesses 
many times the coherence length\cite{Bozovic:2004p506}.
Other experiments on bilayers have shown transition temperatures higher than that of
either layer in isolation\cite{Yuli:2008p507,Berg:2008p508,Gozar:2008p509}.
Microscopic theories to explain these phenomena include Josephson
coupling between superconducting islands\cite{Kresin:2006p830},
and enhancement of the phase stiffness in underdoped regions
by proximity to overdoped regions\cite{Berg:2008p508,PhysRevLett.101.156401,PhysRevB.79.174509}.
Such scenarios are inherently interesting due to the possibility that
interface superconductivity can occur at temperatures above the maximum
possible in bulk samples. To date no experiments have been performed to
investigate proximity effects locally at microscopic length scales.
Here we report a proximity effect in the cuprate superconductor \bscco\ using
scanning tunneling microscopy (STM) and the intrinsic nanoscale spatial
variation of the sample. We show that indeed patches of the sample with
weaker superconductivity can be enhanced if surrounded by patches with
stronger superconductivity, demonstrating that the collapse of superconductivity
is caused by more than local thermal pair breaking.

\begin{figure}
\includegraphics[width=75mm]{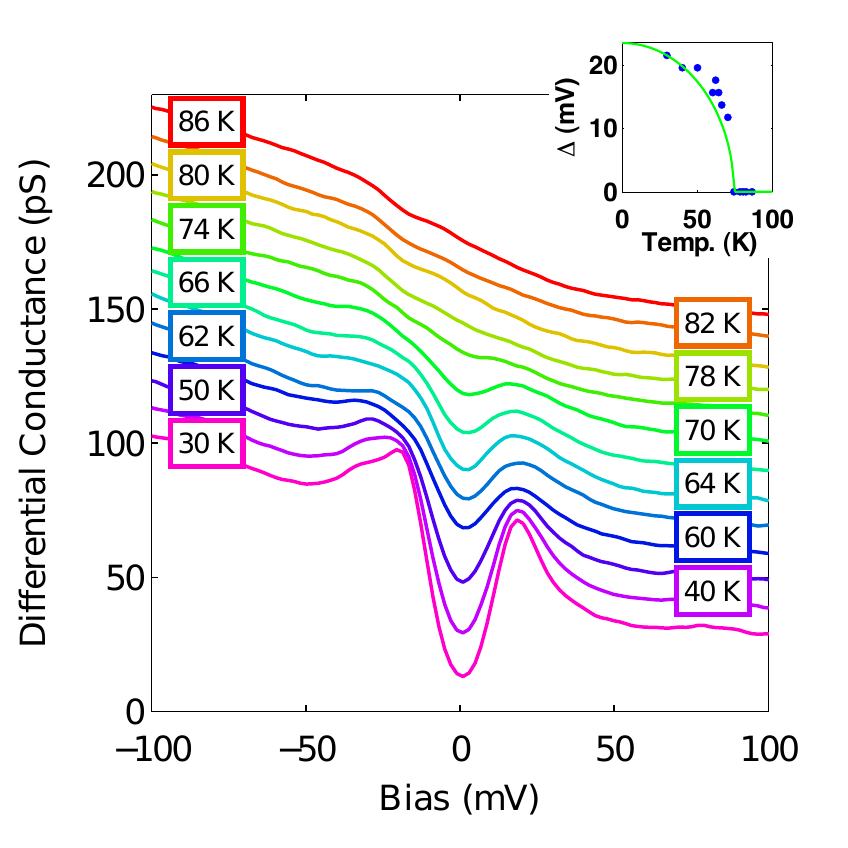}
\caption{\label{fig:spec}Representative $dI/dV$ spectrum at the same point at different
temperatures. The setpoint was at -200~mV. The evolution of the spectrum shows a
gap that closes near 74~K ($T_c$ is 65~K). The inset shows the gap as extracted from the peak location as a function of temperature.}
\end{figure}

Our experiments were performed in a compact variable temperature scanning tunneling microscope (STM), with the unique ability to acquire spectroscopy on nearly identical grids of points over a wide range of temperatures.
This allows us to perform spatial correlations using data from multiple temperatures. The pairing gap is measured from the local differential conductance ($dI/dV$), which is proportional to the local density of states.
Figure \ref{fig:spec} shows a representative $dI/dV$ spectrum for the high temperature superconductor \bscco (Bi-2212). At 30~K there is a visible gap in the spectrum, by 70~K it has diminished considerably,
and by 78~K it is gone completely. The gap has been previously described by a $d$-wave
extension to the BCS theory with the addition of a lifetime broadening term\cite{Alldredge:2008p568,Pasupathy:2008p298}.

For high temperature cuprate superconductors, the superconducting gap
varies from smallest to largest by a factor of two on nanometer scales\cite{Gomes:2007p301,Howald:2001p505,Pan:2001p393,McElroy:2005p326}. By identifying the same atoms
in the topography at multiple temperatures we are able to measure dI/dV spectra
for the same point at different temperatures.
Recently, we have used this technique to visualize the development of superconductivity
on the atomic scale in Bi-2212.
We have found that the temperature dependence of the spectroscopic gap ($\Delta(r,T)$)
is inhomogenous, and that the gap vanishes at a local temperature $T_p(r)$,
which is positively correlated in space with the gap size\cite{Gomes:2007p301}.
Furthermore, the spectroscopic gap below $T_c$ is spatially correlated to the density of states at the Fermi level well above $T_c$ where the gap is absent\cite{Pasupathy:2008p298}.
For this work we simplify the analysis of a large number of points by using a simple,
operational definition for the gap and the pairing temperature.
We follow the definition used previously\cite{Howald:2001p505,Gomes:2007p301},
defining the gap value as the point of maximum differential conductance for
positive bias. This simple procedure agrees with more elaborate fitting schemes
to within 10\% in most cases. We also need a simple method to describe
the temperature evolution of the gap. For this, we define the pairing temperature $T_p$,
which is the highest temperature at which a maximum in the $dI/dV$ spectrum can
be observed within experimental resolution at positive bias.

In underdoped samples, multiple energy features are present in STM spectroscopy\cite{Gomes:2007p301}. While high energy features survive to temperatures well above $T_c$, lower energy features show the closing of a nodal gap near $T_c$\cite{Boyer:2007p633,Pushp:2009p297}. 
However, for strongly overdoped samples pseudogap behavior is absent, there is only one spectroscopic feature in tunneling data, and the median value of $T_p$ is near the resistive transition temperature $T_c$. Furthermore, an inhomogeneous superconducting transition is supported by Nernst measurements\cite{Wang:2006p634} and
$\mu$SR\cite{Sonier:2008p637}. 
Therefore, $\Delta$ may be taken as the superconducting gap, and $T_p$ as the temperature
for the onset of pairing. We have done detailed measurements of the
local $T_p$ and the local $\Delta$ for a grid of points on an overdoped sample of Bi-2212 with
$T_c = 65\textrm{ K}$, for a range of temperatures from 50~K to 76~K.
These measurements provided evidence that nanoscale regions distinguished
by inhomogeneity can influence each other via the proximity effect.
We show that the fractional variation in $T_p$ is less than that for $\Delta$.
The pairing order parameter in a disordered sample
has been analyzed theoretically at low temperature\cite{fang:017007,balatsky:373} and near the
transition\cite{andersen:060501}, but a direct comparison
of spatial variation between $T_p$ and $\Delta$ was not made.
Our measurements show that the length scale for correlations in $T_p$ is longer by 20\%
than the length scale for the gap. Furthermore, we find that the local gap
can disappear at distinct temperatures for different spatial positions
on the sample, even if these positions have the same magnitude of gap at
low temperature, and this anomalous temperature evolution of the local gap
is influenced by the size of gaps in the surrounding 1 nm region.

\begin{figure*}
\includegraphics[width=150mm]{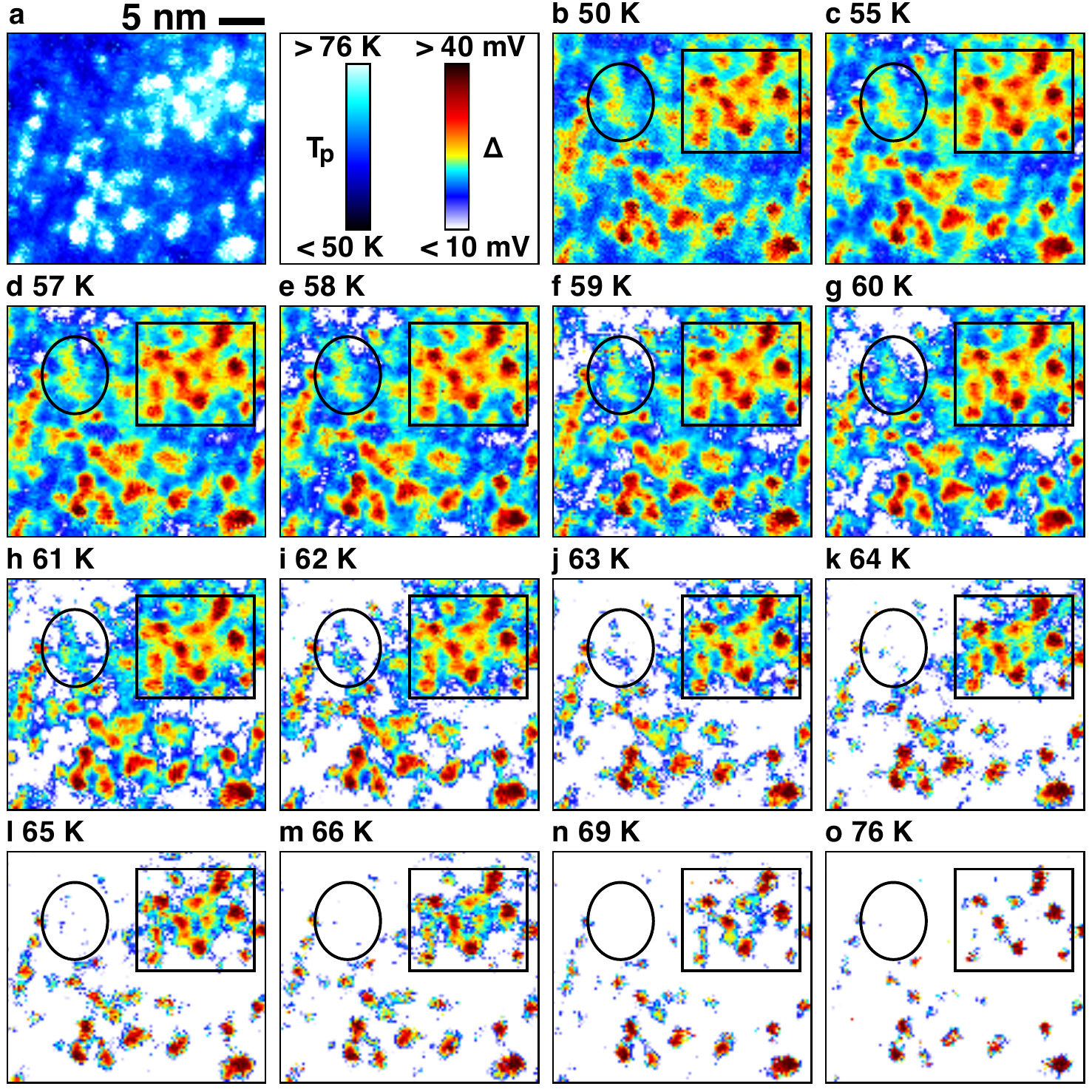}
\caption{\label{fig:manymaps}Evolution of the gap distribution with temperature.
\textbf{a}, map of the local pairing temperature ($T_p$). The pairing temperature
is determined from the sequence of images in \textbf{b}-\textbf{o} by finding the
temperature above which there is no maximum in the conductance on the
positive bias side. \textbf{b}-\textbf{o}, a 25 x 28 nm ``gap map'' taken at several different temperatures
showing the distribution of gaps in real space. At each temperature, the
gap map has been aligned to allow point-by-point comparison between maps.}
\end{figure*}

\begin{figure}
\includegraphics[width=75mm]{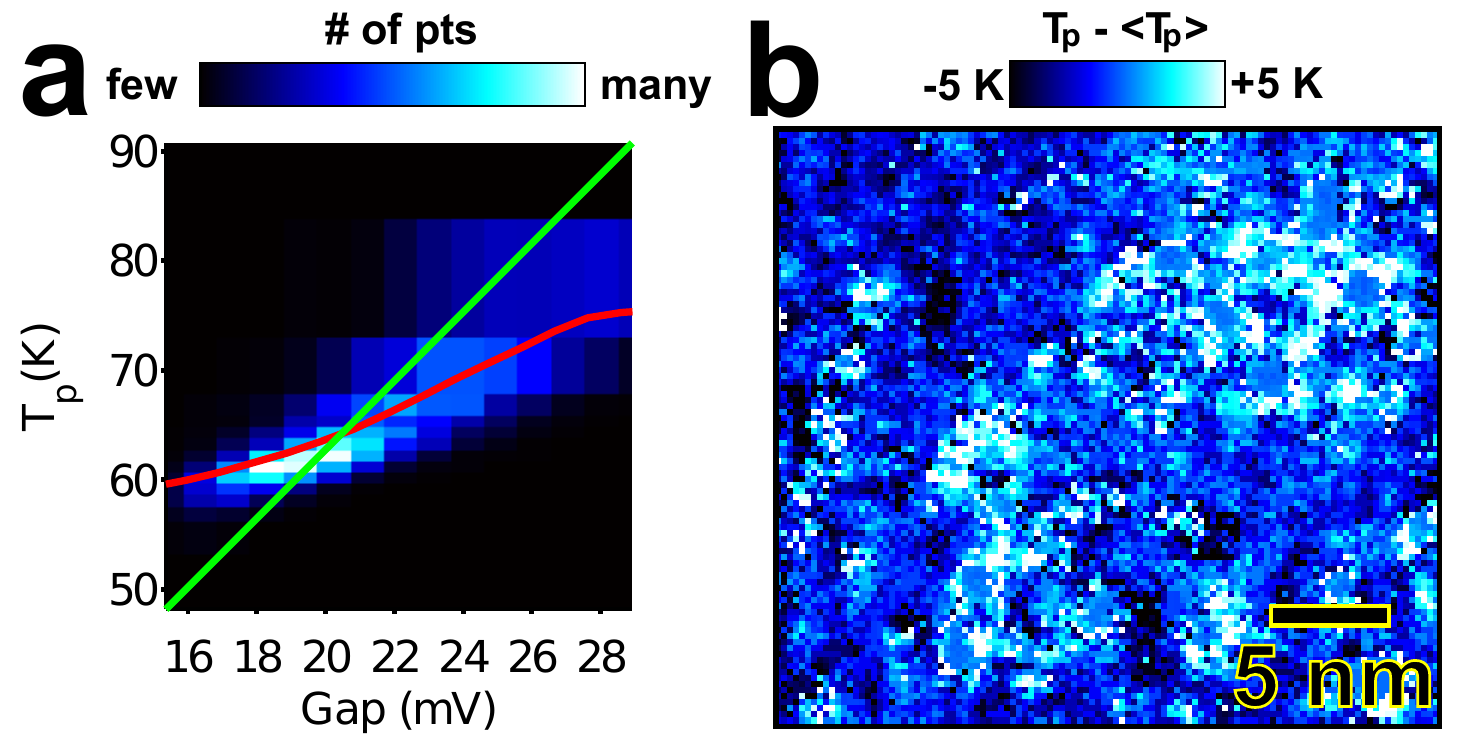}
\caption{\label{fig:hist}Relation between $\Delta$ and $T_p$. \textbf{a}, two dimensional histogram of $\Delta$ and $T_p$ values, showing the strong correlation. The green line represents the linear relation $2\Delta/k_BT_p = 7.4$, which is determined from the median values of $\Delta$ and $T_p$. The red line represents the average value of $T_p$ as a function of $\Delta$. \textbf{b}, The deviation in $T_p$ from the average, that is $\delta T_p = T_p - \left<T_p \right>$, where $\left<T_p \right>$ is the average $T_p$ for gaps with the same size.}
\end{figure}

We have the spectral evolution for more than 10,000 points in a 25 x 28 nm
grid\footnote{These data are taken in the same area of the same sample as used
in Ref.~\onlinecite{Pasupathy:2008p298}} at fourteen different temperatures from 50~K to 76~K.
For each temperature,
we can determine the gap at each point, and plot a series of ``gap maps'',
or 2D false color images showing the spatial distribution of gaps,
shown in figure \ref{fig:manymaps}b-o. Using simultaneously recorded topographic information,
we can align gap maps taken at different temperatures, despite their being taken
over slightly different areas with varying amounts of thermal drift and piezo creep.
This is accomplished by interpolating the two dimensional data in order to maximize the agreement between temperatures. To avoid biasing the results all alignment is based exclusively on topographic data.
Points with median-sized gaps surrounded by large gapped regions (consider points with gaps near 20~mV within the box in the upper right of the panels of figure \ref{fig:manymaps}) survive to higher temperatures
than regions with similar absolute gap sizes elsewhere (consider the oval shaped region in the upper left), which supports our main claim.
From the data of figure \ref{fig:manymaps}b-o, we can determine $T_p$ for each point,
restricted to the temperature range for which we have data,
so that we can test our claims statistically. Figure \ref{fig:manymaps}a
shows the spatial distribution of $T_p$. This is well correlated to the low temperature gap value
from figure \ref{fig:manymaps}b, as we expect if the temperature evolution is determined locally.
Figure \ref{fig:hist}a shows a histogram of all measured gaps and their
corresponding $T_p$ values, confirming that the two quantities are well
correlated (81\%). If we use the median value of the gap (20~mV) and the median value
of $T_p$ (64~K), we obtain $2\Delta/k_BT_p = 7.4$ (the green line in figure \ref{fig:hist}a),
consistent with previous work\cite{Gomes:2007p301}.
By choosing bins of gap size and averaging the $T_p$ values in each bin,
we can establish a more detailed relation between $\Delta$ and $T_p$
(the red line in figure \ref{fig:hist}a). The slope of the distribution is significantly
shallower than the linear relation $2\Delta/k_BT_p = 7.4$.
That is, the relative variation of $\Delta$ is greater than that for $T_p$. Indeed, this is quite a significant effect, as it means that, for example, 17~mV gaps will survive, on average, to temperatures 6~K above where one would expect based on a linear relation.
We take this as evidence that $T_p$ is determined not only by the local gap,
but also by the gap in the surrounding region.
Therefore, the variation in $T_p$ will be reduced,
as this quantity is spatially averaged over some window.

We have found direct evidence that the temperature evolution can be
more accurately predicted given knowledge of the gap in the vicinity of a point,
compared with knowledge of the gap at a single point only.
One can clearly see from figure \ref{fig:hist}a that knowledge of the gap
can only determine $T_p$ in a statistical way, that is, different points of equal gap size have measurably different values of $T_p$. If the pairing temperature were to depend only
locally on the gap, the distribution of $T_p$ values with the same gap
would only be due to random fluctuations. Therefore, when we subtract
the mean $T_p$ value (based on gap size) from each measured $T_p$ value,
the result should be uncorrelated noise. However, figure \ref{fig:hist}b shows a
significant spatial pattern associated with this quantity, which we call $\delta T_p$.

\begin{figure}
\includegraphics[width=75mm]{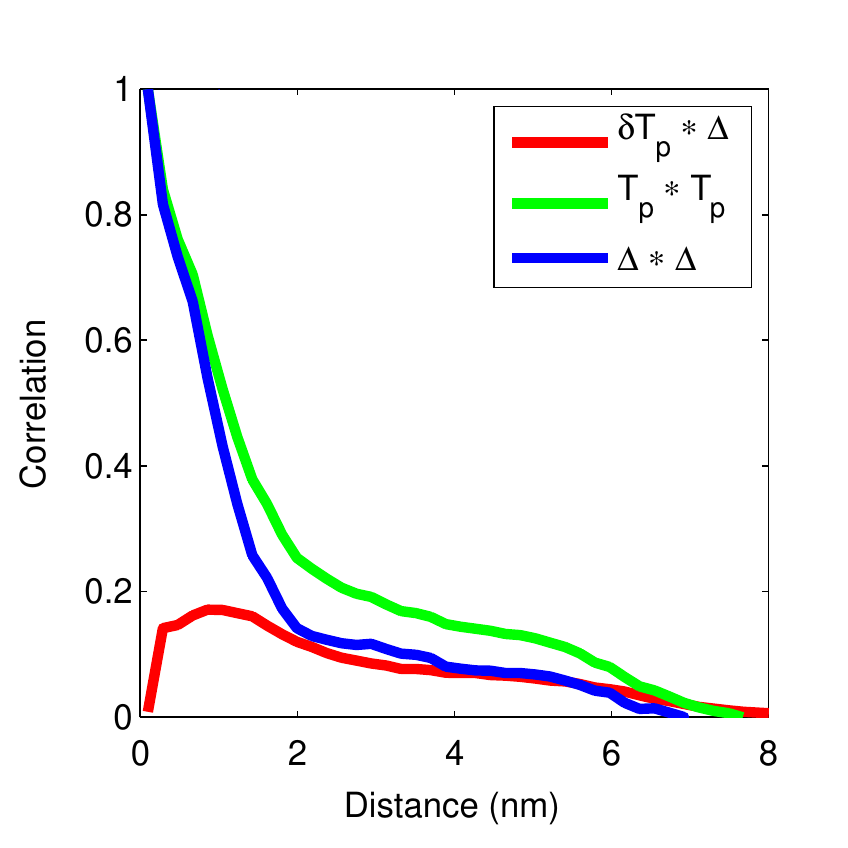}
\caption{\label{fig:corr}Correlation length scales. The cross correlation between $\delta T_p$ and $\Delta$ as a function of separation, and the autocorrelations of $\Delta$ and $T_p$, radially averaged. The cross correlation shows that areas with large $T_p$ values for their gap size are on average surrounded by larger gap regions. The autocorrelations show that $T_p$ has stronger correlation on length scales above 1~nm. The results are not significantly different if averaged along the $a$ or $b$ axis instead of radially.}
\end{figure}

We can understand the meaning of $\delta T_p$ and relate it to the
gap map using the cross correlation as a function of separation distance.
We define the cross correlation as
\begin{eqnarray}
\label{eqn:cross_corr}\rho(\mathbf{x}) = \frac{1}{N\sigma_{\delta T_p}\sigma_\Delta}\sum_{\mathbf{y}} \delta T_p(\mathbf{y})\Delta(\mathbf{y}-\mathbf{x}),
\end{eqnarray}
where $\sigma_{\delta T_p}$ and $\sigma_\Delta$ are the standard deviations of $\delta T_p$ and $\Delta$ respectively, and $\mathbf{y}$ runs over the $N$ available data points. Although the supermodulation on the $b$ axis does have a small correlation 
with the gap, in practice $\rho$ is isotropic within the noise, so we plot a circular average in figure \ref{fig:corr}.
For zero separation, the cross correlations mathematically must be zero,
because we have explicitly eliminated the local relation between $\Delta$ and $T_p$
when we subtract the mean to produce $\delta T_p$. The peak at positive separation
indicates that $T_p$ at a particular point is significantly determined by gaps 1~nm away,
which we attribute to the proximity effect. In addition, there is a long tail of positive correlation,
indicating that $T_p$ is also dependent on gaps several nanometers away.
Figure \ref{fig:corr} also shows the autocorrelation of $\Delta$ and $T_p$ averaged
along both axes and plotted as a function of separation, showing that
the pairing temperature has a longer length scale (1.1~nm at half-maximum)
than the gap (0.9~nm at half-maximum). These correlations occur on length scales longer enough than the 0.25~nm pixel spacing to rule out ``smoothing'' due to the interpolation used in aligning between different temperatures. The strength of $T_p$ correlations
is also higher than that of gap correlations over the tail of the distribution (1 -- 6~nm).
Although the exact shape of this tail may reflect details of the specific area chosen for study,
the fact that the $T_p$ correlation is consistently higher and the $\delta T_p$ -- $\Delta$ cross correlation also extends to this distance indicates that the process by which
the gap collapses is more sensitive to the surrounding several nanometers than the
process that sets the gap energy scale.

We have performed the first detailed measurements of the temperature
evolution of the superconducting gap for a large number of points in
overdoped \bscco. Due to dopants and crystalline disorder, the tendency toward pairing is natrually inhomogeneous. Correspondingly, the superconducting gap
is also inhomogeneous, albeit smoothed out by interactions between regions. The temperature evolution is also inhomogeneous, but we have found that the compared to the gap the magnitude of its variation is more and the length scale of its variation is shorter, that is, the temperature evolution is even more smoothed out. Also, regions of the sample with a small gap magnitude can have this gap survive to
higher temperatures if surrounded by larger gapped regions. We speculate
that the closing of the gap may have to do not just with thermal pair-breaking,
but also with phase fluctuations, so that regions bordered by larger gapped
regions have the phase pinned up to a higher temperature than similar regions
bordered by small gapped regions.


\begin{thebibliography}{20}
\expandafter\ifx\csname natexlab\endcsname\relax\def\natexlab#1{#1}\fi
\expandafter\ifx\csname bibnamefont\endcsname\relax
  \def\bibnamefont#1{#1}\fi
\expandafter\ifx\csname bibfnamefont\endcsname\relax
  \def\bibfnamefont#1{#1}\fi
\expandafter\ifx\csname citenamefont\endcsname\relax
  \def\citenamefont#1{#1}\fi
\expandafter\ifx\csname url\endcsname\relax
  \def\url#1{\texttt{#1}}\fi
\expandafter\ifx\csname urlprefix\endcsname\relax\def\urlprefix{URL }\fi
\providecommand{\bibinfo}[2]{#2}
\providecommand{\eprint}[2][]{\url{#2}}

\bibitem[{\citenamefont{Gomes et~al.}(2007)\citenamefont{Gomes, Pasupathy,
  Pushp, Ono, Ando, and Yazdani}}]{Gomes:2007p301}
\bibinfo{author}{\bibfnamefont{K.~K.} \bibnamefont{Gomes}},
  \bibinfo{author}{\bibfnamefont{A.~N.} \bibnamefont{Pasupathy}},
  \bibinfo{author}{\bibfnamefont{A.}~\bibnamefont{Pushp}},
  \bibinfo{author}{\bibfnamefont{S.}~\bibnamefont{Ono}},
  \bibinfo{author}{\bibfnamefont{Y.}~\bibnamefont{Ando}}, \bibnamefont{and}
  \bibinfo{author}{\bibfnamefont{A.}~\bibnamefont{Yazdani}},
  \bibinfo{journal}{Nature} \textbf{\bibinfo{volume}{447}},
  \bibinfo{pages}{569} (\bibinfo{year}{2007}).

\bibitem[{\citenamefont{Howald et~al.}(2001)\citenamefont{Howald, Fournier, and
  Kapitulnik}}]{Howald:2001p505}
\bibinfo{author}{\bibfnamefont{C.}~\bibnamefont{Howald}},
  \bibinfo{author}{\bibfnamefont{P.}~\bibnamefont{Fournier}}, \bibnamefont{and}
  \bibinfo{author}{\bibfnamefont{A.}~\bibnamefont{Kapitulnik}},
  \bibinfo{journal}{Physical Review B} \textbf{\bibinfo{volume}{64}},
  \bibinfo{pages}{100504(R)} (\bibinfo{year}{2001}).

\bibitem[{\citenamefont{Pan et~al.}(2001)\citenamefont{Pan, O'Neal, Badzey,
  Chamon, Ding, Engelbrecht, Wang, Eisaki, Uchida, Gupta
  et~al.}}]{Pan:2001p393}
\bibinfo{author}{\bibfnamefont{S.~H.} \bibnamefont{Pan}},
  \bibinfo{author}{\bibfnamefont{J.~P.} \bibnamefont{O'Neal}},
  \bibinfo{author}{\bibfnamefont{R.~L.} \bibnamefont{Badzey}},
  \bibinfo{author}{\bibfnamefont{C.}~\bibnamefont{Chamon}},
  \bibinfo{author}{\bibfnamefont{H.}~\bibnamefont{Ding}},
  \bibinfo{author}{\bibfnamefont{J.~R.} \bibnamefont{Engelbrecht}},
  \bibinfo{author}{\bibfnamefont{Z.}~\bibnamefont{Wang}},
  \bibinfo{author}{\bibfnamefont{H.}~\bibnamefont{Eisaki}},
  \bibinfo{author}{\bibfnamefont{S.}~\bibnamefont{Uchida}},
  \bibinfo{author}{\bibfnamefont{A.~K.} \bibnamefont{Gupta}},
  \bibnamefont{et~al.}, \bibinfo{journal}{Nature}
  \textbf{\bibinfo{volume}{413}}, \bibinfo{pages}{282} (\bibinfo{year}{2001}).

\bibitem[{\citenamefont{McElroy et~al.}(2005)\citenamefont{McElroy, Lee,
  Slezak, Lee, Eisaki, Uchida, and Davis}}]{McElroy:2005p326}
\bibinfo{author}{\bibfnamefont{K.}~\bibnamefont{McElroy}},
  \bibinfo{author}{\bibfnamefont{J.}~\bibnamefont{Lee}},
  \bibinfo{author}{\bibfnamefont{J.}~\bibnamefont{Slezak}},
  \bibinfo{author}{\bibfnamefont{D.}~\bibnamefont{Lee}},
  \bibinfo{author}{\bibfnamefont{H.}~\bibnamefont{Eisaki}},
  \bibinfo{author}{\bibfnamefont{S.}~\bibnamefont{Uchida}}, \bibnamefont{and}
  \bibinfo{author}{\bibfnamefont{J.}~\bibnamefont{Davis}},
  \bibinfo{journal}{Science} \textbf{\bibinfo{volume}{309}},
  \bibinfo{pages}{1048} (\bibinfo{year}{2005}).

\bibitem[{\citenamefont{Bozovic et~al.}(2004)\citenamefont{Bozovic, Logvenov,
  Verhoeven, Caputo, Goldobin, and Beasley}}]{Bozovic:2004p506}
\bibinfo{author}{\bibfnamefont{I.}~\bibnamefont{Bozovic}},
  \bibinfo{author}{\bibfnamefont{G.}~\bibnamefont{Logvenov}},
  \bibinfo{author}{\bibfnamefont{M.~A.~J.}~\bibnamefont{Verhoeven}},
  \bibinfo{author}{\bibfnamefont{P.}~\bibnamefont{Caputo}},
  \bibinfo{author}{\bibfnamefont{E.}~\bibnamefont{Goldobin}}, \bibnamefont{and}
  \bibinfo{author}{\bibfnamefont{M.~R.}~\bibnamefont{Beasley}},
  \bibinfo{journal}{Physical Review Letters} \textbf{\bibinfo{volume}{93}},
  \bibinfo{pages}{157002} (\bibinfo{year}{2004}).

\bibitem[{\citenamefont{Yuli et~al.}(2008)\citenamefont{Yuli, Asulin, Millo,
  Orgad, Iomin, and Koren}}]{Yuli:2008p507}
\bibinfo{author}{\bibfnamefont{O.}~\bibnamefont{Yuli}},
  \bibinfo{author}{\bibfnamefont{I.}~\bibnamefont{Asulin}},
  \bibinfo{author}{\bibfnamefont{O.}~\bibnamefont{Millo}},
  \bibinfo{author}{\bibfnamefont{D.}~\bibnamefont{Orgad}},
  \bibinfo{author}{\bibfnamefont{L.}~\bibnamefont{Iomin}}, \bibnamefont{and}
  \bibinfo{author}{\bibfnamefont{G.}~\bibnamefont{Koren}},
  \bibinfo{journal}{Physical Review Letters} \textbf{\bibinfo{volume}{101}},
  \bibinfo{pages}{057005} (\bibinfo{year}{2008}).

\bibitem[{\citenamefont{Berg et~al.}(2008)\citenamefont{Berg, Orgad, and
  Kivelson}}]{Berg:2008p508}
\bibinfo{author}{\bibfnamefont{E.}~\bibnamefont{Berg}},
  \bibinfo{author}{\bibfnamefont{D.}~\bibnamefont{Orgad}}, \bibnamefont{and}
  \bibinfo{author}{\bibfnamefont{S.~A.}~\bibnamefont{Kivelson}},
  \bibinfo{journal}{Physical Review B} \textbf{\bibinfo{volume}{78}},
  \bibinfo{pages}{094509} (\bibinfo{year}{2008}).

\bibitem[{\citenamefont{Gozar et~al.}(2008)\citenamefont{Gozar, Logvenov,
  Kourkoutis, Bollinger, Giannuzzi, Muller, and Bozovic}}]{Gozar:2008p509}
\bibinfo{author}{\bibfnamefont{A.}~\bibnamefont{Gozar}},
  \bibinfo{author}{\bibfnamefont{G.}~\bibnamefont{Logvenov}},
  \bibinfo{author}{\bibfnamefont{L.~F.} \bibnamefont{Kourkoutis}},
  \bibinfo{author}{\bibfnamefont{A.~T.} \bibnamefont{Bollinger}},
  \bibinfo{author}{\bibfnamefont{L.~A.} \bibnamefont{Giannuzzi}},
  \bibinfo{author}{\bibfnamefont{D.~A.} \bibnamefont{Muller}},
  \bibnamefont{and} \bibinfo{author}{\bibfnamefont{I.}~\bibnamefont{Bozovic}},
  \bibinfo{journal}{Nature} \textbf{\bibinfo{volume}{455}},
  \bibinfo{pages}{782} (\bibinfo{year}{2008}).

\bibitem[{\citenamefont{Kresin et~al.}(2006)\citenamefont{Kresin, Ovchinnikov,
  and Wolf}}]{Kresin:2006p830}
\bibinfo{author}{\bibfnamefont{V.}~\bibnamefont{Kresin}},
  \bibinfo{author}{\bibfnamefont{Y.}~\bibnamefont{Ovchinnikov}},
  \bibnamefont{and} \bibinfo{author}{\bibfnamefont{S.}~\bibnamefont{Wolf}},
  \bibinfo{journal}{Physics Reports} \textbf{\bibinfo{volume}{431}},
  \bibinfo{pages}{231} (\bibinfo{year}{2006}).

\bibitem[{\citenamefont{Okamoto and Maier}(2008)}]{PhysRevLett.101.156401}
\bibinfo{author}{\bibfnamefont{S.}~\bibnamefont{Okamoto}} \bibnamefont{and}
  \bibinfo{author}{\bibfnamefont{T.~A.} \bibnamefont{Maier}},
  \bibinfo{journal}{Phys. Rev. Lett.} \textbf{\bibinfo{volume}{101}},
  \bibinfo{pages}{156401} (\bibinfo{year}{2008}).

\bibitem[{\citenamefont{Goren and Altman}(2009)}]{PhysRevB.79.174509}
\bibinfo{author}{\bibfnamefont{L.}~\bibnamefont{Goren}} \bibnamefont{and}
  \bibinfo{author}{\bibfnamefont{E.}~\bibnamefont{Altman}},
  \bibinfo{journal}{Phys. Rev. B} \textbf{\bibinfo{volume}{79}},
  \bibinfo{pages}{174509} (\bibinfo{year}{2009}).

\bibitem[{\citenamefont{Pasupathy et~al.}(2008)\citenamefont{Pasupathy, Pushp,
  Gomes, Parker, Wen, Xu, Gu, Ono, Ando, and Yazdani}}]{Pasupathy:2008p298}
\bibinfo{author}{\bibfnamefont{A.}~\bibnamefont{Pasupathy}},
  \bibinfo{author}{\bibfnamefont{A.}~\bibnamefont{Pushp}},
  \bibinfo{author}{\bibfnamefont{K.}~\bibnamefont{Gomes}},
  \bibinfo{author}{\bibfnamefont{C.}~\bibnamefont{Parker}},
  \bibinfo{author}{\bibfnamefont{J.}~\bibnamefont{Wen}},
  \bibinfo{author}{\bibfnamefont{Z.}~\bibnamefont{Xu}},
  \bibinfo{author}{\bibfnamefont{G.}~\bibnamefont{Gu}},
  \bibinfo{author}{\bibfnamefont{S.}~\bibnamefont{Ono}},
  \bibinfo{author}{\bibfnamefont{Y.}~\bibnamefont{Ando}}, \bibnamefont{and}
  \bibinfo{author}{\bibfnamefont{A.}~\bibnamefont{Yazdani}},
  \bibinfo{journal}{Science} \textbf{\bibinfo{volume}{320}},
  \bibinfo{pages}{196} (\bibinfo{year}{2008}).

\bibitem[{\citenamefont{Boyer et~al.}(2007)\citenamefont{Boyer, Wise,
  Chatterjee, Yi, Kondo, Takeuchi, Ikuta, and Hudson}}]{Boyer:2007p633}
\bibinfo{author}{\bibfnamefont{M.~C.} \bibnamefont{Boyer}},
  \bibinfo{author}{\bibfnamefont{W.~D.} \bibnamefont{Wise}},
  \bibinfo{author}{\bibfnamefont{K.}~\bibnamefont{Chatterjee}},
  \bibinfo{author}{\bibfnamefont{M.}~\bibnamefont{Yi}},
  \bibinfo{author}{\bibfnamefont{T.}~\bibnamefont{Kondo}},
  \bibinfo{author}{\bibfnamefont{T.}~\bibnamefont{Takeuchi}},
  \bibinfo{author}{\bibfnamefont{H.}~\bibnamefont{Ikuta}}, \bibnamefont{and}
  \bibinfo{author}{\bibfnamefont{E.~W.} \bibnamefont{Hudson}},
  \bibinfo{journal}{Nature Physics} \textbf{\bibinfo{volume}{3}},
  \bibinfo{pages}{802} (\bibinfo{year}{2007}).

\bibitem[{\citenamefont{Pushp et~al.}(2009)\citenamefont{Pushp, Parker,
  Pasupathy, Gomes, Ono, Wen, Xu, Gu, and Yazdani}}]{Pushp:2009p297}
\bibinfo{author}{\bibfnamefont{A.}~\bibnamefont{Pushp}},
  \bibinfo{author}{\bibfnamefont{C.}~\bibnamefont{Parker}},
  \bibinfo{author}{\bibfnamefont{A.}~\bibnamefont{Pasupathy}},
  \bibinfo{author}{\bibfnamefont{K.}~\bibnamefont{Gomes}},
  \bibinfo{author}{\bibfnamefont{S.}~\bibnamefont{Ono}},
  \bibinfo{author}{\bibfnamefont{J.}~\bibnamefont{Wen}},
  \bibinfo{author}{\bibfnamefont{Z.}~\bibnamefont{Xu}},
  \bibinfo{author}{\bibfnamefont{G.}~\bibnamefont{Gu}}, \bibnamefont{and}
  \bibinfo{author}{\bibfnamefont{A.}~\bibnamefont{Yazdani}},
  \bibinfo{journal}{Science} \textbf{\bibinfo{volume}{324}},
  \bibinfo{pages}{1689} (\bibinfo{year}{2009}).

\bibitem[{\citenamefont{Wang et~al.}(2006)\citenamefont{Wang, Li, and
  Ong}}]{Wang:2006p634}
\bibinfo{author}{\bibfnamefont{Y.}~\bibnamefont{Wang}},
  \bibinfo{author}{\bibfnamefont{L.}~\bibnamefont{Li}}, \bibnamefont{and}
  \bibinfo{author}{\bibfnamefont{N.~P.}~\bibnamefont{Ong}},
  \bibinfo{journal}{Physical Review B} \textbf{\bibinfo{volume}{73}},
  \bibinfo{pages}{024510} (\bibinfo{year}{2006}).

\bibitem[{\citenamefont{Sonier et~al.}(2008)\citenamefont{Sonier, Ilton,
  Pacradouni, Kaiser, Sabok-Sayr, Ando, Komiya, Hardy, Bonn, Liang, and Atkinson}}]{Sonier:2008p637}
\bibinfo{author}{\bibfnamefont{J.~E.} \bibnamefont{Sonier}},
  \bibinfo{author}{\bibfnamefont{M.}~\bibnamefont{Ilton}},
  \bibinfo{author}{\bibfnamefont{V.}~\bibnamefont{Pacradouni}},
  \bibinfo{author}{\bibfnamefont{C.~V.} \bibnamefont{Kaiser}},
  \bibinfo{author}{\bibfnamefont{S.~A.} \bibnamefont{Sabok-Sayr}},
  \bibinfo{author}{\bibfnamefont{Y.}~\bibnamefont{Ando}},
  \bibinfo{author}{\bibfnamefont{S.}~\bibnamefont{Komiya}},
  \bibinfo{author}{\bibfnamefont{W.~N.} \bibnamefont{Hardy}},
  \bibinfo{author}{\bibfnamefont{D.~A.} \bibnamefont{Bonn}},
  \bibinfo{author}{\bibfnamefont{R.}~\bibnamefont{Liang}}, \bibnamefont{and}
    \bibinfo{author}{\bibfnamefont{W.A}~\bibnamefont{Atkinson}},
    \bibinfo{journal}{Physical Review Letters}
  \textbf{\bibinfo{volume}{101}}, \bibinfo{pages}{117001}
  (\bibinfo{year}{2008}).

\bibitem[{\citenamefont{Fang et~al.}(2006)\citenamefont{Fang, Capriotti,
  Scalapino, Kivelson, Kaneko, Greven, and Kapitulnik}}]{fang:017007}
\bibinfo{author}{\bibfnamefont{A.~C.} \bibnamefont{Fang}},
  \bibinfo{author}{\bibfnamefont{L.}~\bibnamefont{Capriotti}},
  \bibinfo{author}{\bibfnamefont{D.~J.} \bibnamefont{Scalapino}},
  \bibinfo{author}{\bibfnamefont{S.~A.} \bibnamefont{Kivelson}},
  \bibinfo{author}{\bibfnamefont{N.}~\bibnamefont{Kaneko}},
  \bibinfo{author}{\bibfnamefont{M.}~\bibnamefont{Greven}}, \bibnamefont{and}
  \bibinfo{author}{\bibfnamefont{A.}~\bibnamefont{Kapitulnik}},
  \bibinfo{journal}{Physical Review Letters} \textbf{\bibinfo{volume}{96}},
  \bibinfo{pages}{017007} (\bibinfo{year}{2006}).

\bibitem[{\citenamefont{Balatsky et~al.}(2006)\citenamefont{Balatsky, Vekhter,
  and Zhu}}]{balatsky:373}
\bibinfo{author}{\bibfnamefont{A.~V.} \bibnamefont{Balatsky}},
  \bibinfo{author}{\bibfnamefont{I.}~\bibnamefont{Vekhter}}, \bibnamefont{and}
  \bibinfo{author}{\bibfnamefont{J.-X.} \bibnamefont{Zhu}},
  \bibinfo{journal}{Reviews of Modern Physics} \textbf{\bibinfo{volume}{78}},
  \bibinfo{pages}{373} (\bibinfo{year}{2006}).

\bibitem[{\citenamefont{Andersen et~al.}(2006)\citenamefont{Andersen, Melikyan,
  Nunner, and Hirschfeld}}]{andersen:060501}
\bibinfo{author}{\bibfnamefont{B.~M.} \bibnamefont{Andersen}},
  \bibinfo{author}{\bibfnamefont{A.}~\bibnamefont{Melikyan}},
  \bibinfo{author}{\bibfnamefont{T.~S.} \bibnamefont{Nunner}},
  \bibnamefont{and} \bibinfo{author}{\bibfnamefont{P.~J.}
  \bibnamefont{Hirschfeld}}, \bibinfo{journal}{Physical Review B}
  \textbf{\bibinfo{volume}{74}}, \bibinfo{pages}{060501(R)}
  (\bibinfo{year}{2006}).

\bibitem[{\citenamefont{Alldredge et~al.}(2008)\citenamefont{Alldredge, Lee,
  McElroy, Wang, Fujita, Kohsaka, Taylor, Eisaki, Uchida, Hirschfeld
  et~al.}}]{Alldredge:2008p568}
\bibinfo{author}{\bibfnamefont{J.~W.} \bibnamefont{Alldredge}},
  \bibinfo{author}{\bibfnamefont{J.}~\bibnamefont{Lee}},
  \bibinfo{author}{\bibfnamefont{K.}~\bibnamefont{McElroy}},
  \bibinfo{author}{\bibfnamefont{M.}~\bibnamefont{Wang}},
  \bibinfo{author}{\bibfnamefont{K.}~\bibnamefont{Fujita}},
  \bibinfo{author}{\bibfnamefont{Y.}~\bibnamefont{Kohsaka}},
  \bibinfo{author}{\bibfnamefont{C.}~\bibnamefont{Taylor}},
  \bibinfo{author}{\bibfnamefont{H.}~\bibnamefont{Eisaki}},
  \bibinfo{author}{\bibfnamefont{S.}~\bibnamefont{Uchida}},
  \bibinfo{author}{\bibfnamefont{P.~J.} \bibnamefont{Hirschfeld}},
  \bibnamefont{et~al.}, \bibinfo{journal}{Nature Physics}
  \textbf{\bibinfo{volume}{4}}, \bibinfo{pages}{319} (\bibinfo{year}{2008}).

\end{thebibliography}
\end{document}